\documentstyle[12pt,psfig]{article}          % usual 12pt
\newskip\humongous \humongous=0pt plus 1000pt minus 1000pt

\newif\ifdtup

\def\Im{\mathop{\rm Im}}

\def\pr#1{#1^\prime}
\def\ltap{\;\raisebox{-.4ex}{\rlap{$\sim$}} \raisebox{.4ex}{$<$}\;}

\def\beq{\begin{equation}}
\def\eeq{\end{equation}}

\def\beqn{\begin{eqnarray}}
\def\eeqn{\end{eqnarray}}
\relax

\def\dotx{\dotx{\dot\overline{x}}}

\relax

\catcode`\@=11

\@addtoreset{equation}{section}
\def\theequation{\thesection\arabic{equation}}

\def\@normalsize{\@setsize\normalsize{15pt}\xiipt\@xiipt
\abovedisplayskip 14pt plus3pt minus3pt%
\belowdisplayskip \abovedisplayskip
\abovedisplayshortskip \z@ plus3pt%
\belowdisplayshortskip 7pt plus3.5pt minus0pt}

\def\small{\@setsize\small{13.6pt}\xipt\@xipt
\abovedisplayskip 13pt plus3pt minus3pt%
\belowdisplayskip \abovedisplayskip
\abovedisplayshortskip \z@ plus3pt%
\belowdisplayshortskip 7pt plus3.5pt minus0pt
\def\@listi{\parsep 4.5pt plus 2pt minus 1pt
     \itemsep \parsep
     \topsep 9pt plus 3pt minus 3pt}}

\@twosidetrue
\relax

\catcode`@=12

\catcode`\@=11

\def\section{\@startsection{section}{1}{\z@}{3.5ex plus 1ex minus
   .2ex}{2.3ex plus .2ex}{\large\bf}}

\def\thesection{\arabic{section}.}

\def\appendix{\setcounter{section}{0}
 \def\thesection{APPENDIX \Alph{section}:}
 \def\theequation{\Alph{section}.\arabic{equation}}}
\def\ps@headings{\def\@oddfoot{}\def\@evenfoot{}
\def\@oddhead{\hbox{}\hfill
 \makebox[.5\textwidth]{\raggedright\ignorespaces --\thepage{}--
 \hfill {}}}  %instead of {\rm FERMILAB--Pub--\FERMIPUB}}}
\def\@evenhead{\@oddhead}
\def\subsectionmark##1{\markboth{##1}{}}
}

\ps@headings

\catcode`\@=12

\def\figcap{\section*{Figure Captions\markboth
 {FIGURECAPTIONS}{FIGURECAPTIONS}}\list
 {Fig. \arabic{enumi}:\hfill}{\settowidth\labelwidth{Fig. 999:}
 \leftmargin\labelwidth
 \advance\leftmargin\labelsep\usecounter{enumi}}}
 \relax
\def\tablecap{\section*{Table Captions\markboth
 {TABLECAPTIONS}{TABLECAPTIONS}}\list
 {Table \arabic{enumi}:\hfill}{\settowidth\labelwidth{Table 999:}
 \leftmargin\labelwidth
 \advance\leftmargin\labelsep\usecounter{enumi}}}
 \relax
\def\reflist{\section*{References\markboth
 {REFLIST}{REFLIST}}\list
 {[\arabic{enumi}]\hfill}{\settowidth\labelwidth{[999]}
 \leftmargin\labelwidth
 \advance\leftmargin\labelsep\usecounter{enumi}}}
 \relax

\catcode`\@=11

\def\ps@headings{\def\@oddfoot{}\def\@evenfoot{}
\def\@oddhead{\hbox{}\hfill
 \makebox[.5\textwidth]{\raggedright\ignorespaces --\thepage{}--
 \hfill {}}}    %instead of {\rm FERMILAB--Pub--\FERMIPUB}}}
\def\@evenhead{\@oddhead}
\def\subsectionmark##1{\markboth{##1}{}}
}

\ps@headings

\catcode`\@=12

\relax

\catcode`\@=11
\def\prm{\fam \z@}
\catcode`\@=12

\relax
\def\pl#1#2#3{{\it Phys. Lett. }{\bf #1}(19#2)#3}

\def\prl#1#2#3{{\it Phys. Rev. Lett. }{\bf #1}(19#2)#3}

\def\prep#1#2#3{{\it Phys. Rep. }{\bf #1}(19#2)#3}
\def\pr#1#2#3{{\it Phys. Rev. }{\bf #1}(19#2)#3}
\def\np#1#2#3{{\it Nucl. Phys. }{\bf #1}(19#2)#3}

\relax

\def    \be             {\begin{equation}}
\def    \beq            {\begin{equation}}
\def    \ee             {\end{equation}}
\def    \eeq            {\end{equation}}
\def    \ba             {\begin{eqnarray}}
\def    \beqn           {\begin{eqnarray}}
\def    \ea             {\end{eqnarray}}
\def    \eeqn           {\end{eqnarray}}

\def    \=              {\;=\;}
\def    \frac           #1#2{{#1 \over #2}}

\def \to   {\mbox{$\rightarrow$}}
\newcommand\sss{\scriptscriptstyle}
\newcommand\as{\alpha_{\sss S}}
\newcommand\aren{\alpha_{\rm ren}}
\newcommand\aseff{\bar{\alpha}_{\sss S}}
\newsavebox\tmpfig

\newsavebox\tttfig

\newcommand\ep{\epsilon}

\newcommand{\ask}{\aseff(k_\perp^2)}

\newcommand{\kp}{k_\perp}

\begin{document}
\begin{titlepage}
\nopagebreak
{\flushright{
        \begin{minipage}{4cm}
        CERN-TH/95-150 \\
        IFUM 507/FT \\
        hep-ph/9506317
        \end{minipage}        }

}
\vfill
\begin{center}
{\LARGE { \bf \sc
Infrared Renormalons \\
and Power Suppressed Effects in $e^+e^-$ Jet Events}}
\vfill
\vskip .5cm
{\bf Paolo NASON%
\footnote{On leave of absence from INFN, Milano, Italy}
and
Michael H. SEYMOUR
}
\vskip .3cm
{CERN, TH Division, Geneva, Switzerland} \\
\vskip .6cm
\end{center}
\nopagebreak
\vfill
%\vskip 3cm
\begin{abstract}
We study the effect of infrared renormalons upon
shape variables that are commonly used to determine the strong
coupling constant in $e^+e^-$ annihilation into hadronic jets.
We consider the model of QCD in the limit of large $n_f$.
We find a wide variety of different
behaviours of shape variables with respect to power suppressed
effects induced by infrared renormalons.
In particular, we find that oblateness
is affected by $1/Q$ non--perturbative effects even
away from the two jet region, and the energy--energy
correlation is affected by $1/Q$ non--perturbative effects
for all values of the angle. On the contrary,
variables like thrust, the $c$ parameter, the heavy jet mass,
and others, do not develop any $1/Q$ correction away from the
two jet region at the leading $n_f$ level.
We argue that $1/Q$ corrections will eventually arise at subleading
$n_f$ level,
but that they could maintain an extra $\as(Q)$ suppression.
We conjecture therefore that the leading power correction to shape variables
will have in general the form $\alpha^n_{\rm S}(Q)/Q$,
and it may therefore be possible to classify shape variables according
to the value of $n$.
\end{abstract}
\vskip 1cm
CERN-TH/95-150 \hfill \\
June 1995 \hfill
\vfill
\end{titlepage}
%$\mbox{}$\addtocounter{page}{-1}\thispagestyle{empty}\newpage
%
\section{Introduction}
Tests of QCD carried out at $e^+e^-$ colliders have received
a considerable boost at LEP and LHC \cite{QCDatLEP},
where, because of the large
centre of mass energy, the perturbative character of jet production
becomes quite prevalent. Although jet studies at LEP provide convincing
evidence of the validity of the perturbative approach, the determination
of the strong coupling from jets cannot be considered as solid as other type
of determinations, like those from the $Z$ hadronic width or
deep inelastic scattering experiments~\cite{WebberGlasgow}.
In fact, even at LEP energies, there are substantial power suppressed
effects that are corrected for using Monte Carlo models. The estimate
of the theoretical error associated with these corrections is
very difficult, and inherently model--dependent.
It would be very desirable to acquire some knowledge of power corrections
from theory alone, without the need to resort
to models.

Some sources of power suppressed effects are in fact understood as
originating from factorial growth of the coefficients of the perturbative
expansion arising either from the large momentum region (UV renormalons)
or from the low momentum region (IR renormalons)
of a certain class of feynman graphs (see ref.~\cite{Mueller}
and references therein).
IR renormalons, UV renormalons and instantons are the only known
sources of factorial growth in the perturbative expansion.
Instantons are known to give corrections that are suppressed by a very
high power of the hard scale involved in the process, while
renormalons may give $1/Q$ corrections to certain quantities.
It has been argued in refs. \cite{WebberAlone} and \cite{Webber} that
$1/Q$ power corrections arising from infrared renormalons are
present in certain jet shape variables, and that these corrections
may also be described in a common framework in terms of a ``frozen''
running coupling constant. In ref.~\cite{Akhoury} the strongest
suggestion is made that the $1/Q$ corrections may factorize,
and it may even be possible to describe power corrections
in Drell-Yan pair production and in jet events in a unified
framework. Power corrections to jet shape variables were also considered
in a simplified model in ref.~\cite{ManoharWise}.

In the present work, we actually compute the effect of renormalons
on jet shape variables in QCD in the limit of
large $n_{f}$ \cite{LargeNfLimit}. In this limit the theory is not
asymptotically free, but we will try to infer the properties of the
full theory just by changing the sign of the first coefficient
of the beta function at the end of our calculation. Our attitude is that
QCD is at least as bad as this limit.

The remainder of this paper proceeds as follows. In Section~2 we will give
an introductory description of our calculation, without going
into any technical detail. In fact, the essence of the physical picture
that we develop is contained in this section. In section~3 we give
a description of the full calculation, and deal with the
subtleties associated with canceling the real and virtual divergences by
performing the calculation in dimensional regularization.
In Section~4 we discuss the term (which we call ``Sudakov'' term) that
has a factorized 3--body form, which we identify with the term discussed
in refs.~\cite{Webber} and \cite{Akhoury}. In Section~5 we discuss
the non-factorizable piece. We show explicitly that this term is
different for thrust and for the heavy jet mass, which are quantities that
have the same definitions at the 3--parton level, but differ at the
4--parton level. In Section~6 we discuss the possible effects of subleading
corrections, and what could be expected in the full QCD theory.
Finally, in Section~7 we give our conclusions.

\section{Infrared renormalon in the large \boldmath$n_f$ limit}
\begin{figure}[htb]
\centerline{\psfig{figure=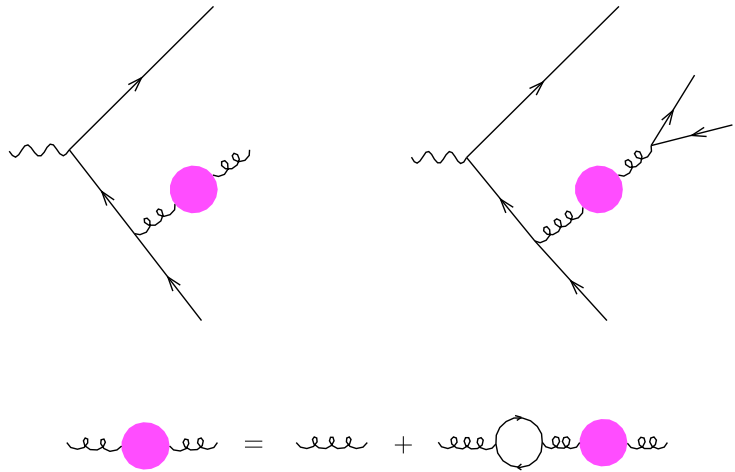,width=10cm,clip=}}
\caption{ \label{threejetsblob}
Dominant diagrams for $e^+e^-$ into jets in the
large $n_f$ limit.
}
\end{figure}

We will first examine the effect of infrared renormalons
in the limit of large $n_f$.
In this case the dominant graphs are those given in
fig.~\ref{threejetsblob}.
%%%%%%%%%%%%%%%%%%%%%%%%
%%%%%%%%%%%%%%%%%%%%%%%%
Notice that there are two types of contribution, one with three partons
in the final state and one with four.
When taking the square of the amplitude with four final state partons
the fermions coming from the gluon splitting
should not interfere with those coming from the photon vertex in order to
give a dominant term in the large $n_f$ limit.

We will now discuss what we expect for the
result of the calculation of jet shape variables.
The aim of this discussion is simply to give the flavour of how
the exact calculation, which will be presented in the following
section, works. All details of the infrared cancelation will be
dealt with in the next section.

In some appropriate renormalization scheme
the inclusion of all vacuum bubbles will amount to the replacement
\beq\label{EvolAlfa}
\as\; \to \;\frac{\as}
{1+\as b_0\log(-k^2/\mu^2)}
\eeq
where
\beq
b_0=-\frac{T_R}{3\pi}\,,
\eeq
and $T_R=n_f/2$.
The contribution of the graphs with three partons in the
final state will be simply given by the Born cross section, with
the replacement of eq.~(\ref{EvolAlfa}) at $k^2=0$. In this
limit our expression vanishes, which is to say, we have no
virtual graphs after we have resummed the whole perturbative
expansion. It does not vanish, however, order by order in perturbation
theory, but has instead an expansion with infrared divergent coefficients.
In the next section we will show that these divergences are canceled
order by order in perturbation theory by the real term.
For the sake of the present illustrative discussion, we will instead
accept the fact that it vanishes, and concentrate on the finite
remainder coming from the real process.

For the corresponding real process, the amplitude will in general
have the form
\beq
A(k^2,\phi)\;d\phi \frac{d\,k^2}{k^2}
\left|\frac{\as}{1+\as\,b_0\left(\log\frac{k^2}{\mu^2}+i\pi\right)} \right|^2
\eeq
where $\phi$ represent here the whole of phase space, except
for the virtuality of the gluon,~$k^2$. In this expression we have
factored out the infrared divergent term $dk^2/k^2$,
so that $A(k^2,\phi)$ is in fact regular for $k^2\to 0$.
Observe that since $k^2$ is positive, the logarithm in the running coupling
acquires an imaginary part. Let us now suppose that we are computing some
infrared safe shape variable $S(k^2,\phi)$.
Infrared safety implies that in the limit of small $k^2$
$S(k^2,\phi)$ goes continuously to its three--body form, so that the
cancelation between real and virtual infrared divergences takes place.
We define
\beq
G(k^2)=\int\; A(k^2,\phi)\; S(k^2,\phi) \;d\,\phi\,.
\eeq
The value of $S$ will be given by
\beq
S=\int \frac{d\,k^2}{k^2}\;G(k^2)\;
\left|\frac{\as}{1+\as\,b_0\left(\log\frac{k^2}{\mu^2}+i\pi\right)} \right|^2
\eeq
which we rewrite as
\beqn
S&=&\int \frac{d\,k^2}{k^2}\;[G(k^2)-G(0)]\;
\frac{\as^2}{\left(1+\as\,b_0\log\frac{k^2}{\mu^2}\right)^2
+\as^2\pi^2 b_0^2}
\nonumber \\ \label{4body}
&& + G(0) \int \frac{d\,k^2}{k^2}\;
\frac{\as^2}{\left(1+\as\,b_0\log\frac{k^2}{\mu^2}\right)^2
+\as^2\pi^2 b_0^2}\,.
\eeqn
For the integral in the second term we get
\beqn &&
G(0)\int \frac{d\,k^2}{k^2}\;
\frac{\as^2}{\left(1+\as\,b_0\log\frac{k^2}{\mu^2}\right)^2
+\as^2\pi^2 b_0^2}
\nonumber\\&&
= \frac{G(0)}{b_0^2\pi}\left[
\arctan\left(\frac{1+\as b_0\log\frac{Q^2}{\mu^2}}{\as b_0\pi}\right)
-\arctan\left(\frac{1+\as b_0\log\frac{\lambda^2}{\mu^2}}{\as b_0\pi}\right)
\right]
\nonumber \\ \label{4bodyresult} &&
= -\frac{G(0)}{b_0^2\pi}\left[
\arctan\left(\frac{\as b_0\pi}{1+\as b_0\log\frac{Q^2}{\mu^2}}\right)
-\arctan\left(\frac{\as b_0\pi}{1+\as b_0\log\frac{\lambda^2}{\mu^2}}\right)
\right]
\eeqn
where $\lambda$ is an infrared cutoff. There is a subtlety in the second
step, where we use the identity $\arctan(x)=\pm\pi/2-\arctan(1/x)$,
in which the $\pi/2$ takes the same sign as $x$.  Thus our substitution
is only correct when the arguments of both arctangents have the same sign.
This condition is violated if $b_0$ is positive, and it might appear that
we have neglected a term proportional to $\pi,$ with no coupling in front.
However, we should remember that we are computing a perturbative expansion,
and that our
algebraic manipulations should always be interpreted as an order by
order expansion in $\as$. We should therefore always be reasoning by
assuming that terms with factors of $\as$ are small, even if
multiplied by infrared or ultraviolet divergent coefficients.
In this sense eq.~(\ref{4bodyresult}) is correct for either sign of $b_0$.

The infrared divergent term we obtain cancels against
the virtual diagram. The cancelation is explicitly shown in the
next section. Here we just assume that it will take place.
In the present context, the real infrared divergent term vanishes
in the same sense in which the virtual term was vanishing.
We see therefore that the only place where we can obtain an infrared
renormalon is the first term of eq.~(\ref{4body}). Let us assume that
$G(k^2)-G(0)\propto k^p$ for small $k,$ and consider the integral
\beqn
I_p(Q^2)&=&\int_0^{Q^2} \frac{d k^2}{k^2}\;\left(\frac{k}{Q}\right)^p\;
\frac{\as^2}{\left(1+\as\,b_0\log\frac{k^2}{\mu^2}\right)^2
+\as^2\pi^2 b_0^2}
\nonumber \\
&=&\int_0^\infty dz\,e^{-pz/2}
\frac{\as^2}{\left(1+\as b_0\log\frac{Q^2}{\mu^2}
-\as b_0 z\right)^2+(\as b_0 \pi)^2}
\nonumber \\
&=&\frac{1}{b_0\pi}\Im\left[ \int_0^\infty dz\,e^{-pz/2}
\frac{\as}{1+\as b_0\log\frac{Q^2}{\mu^2}
-\as b_0 z-i\as b_0 \pi}\right]
\nonumber \\ \label{Ipdef}
&=& \frac{1}{b_0\pi}\Im\left[ \int_{z_0}^{z_0+\infty} dz^\prime\,
e^{-p(z^\prime-z_0)/2}
\frac{\as}{1-\as b_0 z^\prime}\right]\,,
\eeqn
where $z=\log\frac{Q^2}{k^2}$, $z^\prime=z+z_0$ and
$z_0=i\pi-\log\frac{Q^2}{\mu^2}$. This becomes
\beqn
I_p(Q^2)&=&
-\frac{1}{b_0\pi}\Im\left[  \int_{0}^{z_0} dz^\prime\,
e^{-p(z^\prime-z_0)/2}
\frac{\as}{1-\as b_0 z^\prime}\right]
\nonumber \\&&
+ \frac{1}{b_0\pi}\Im \left[ \int_{0}^{\infty} dz^\prime\,
e^{-p(z^\prime-z_0)/2}
\frac{\as}{1-\as b_0 z^\prime-i\ep}\right]
\eeqn
and the first term is analytic in $\as$. Rescaling
$z^\prime$ we finally get
\beqn
I_p(Q^2)&=&
-\frac{1}{b_0\pi}\Im\left[  \int_{0}^{z_0} dz^\prime
e^{-p(z^\prime-z_0)/2}
\frac{\as}{1-\as b_0 z^\prime}\right]
\nonumber \\&&
 +\frac{1}{b_0^2\,\pi}\Im\left[ e^{i\,p\,\pi/2} \left(\frac{\mu}{Q}\right)^p
\int_{0}^{\infty} dz
\frac{\exp(-z/\as)}{\frac{p}{2 b_0}-z-i\ep} \right]\,.
\label{IpDef}
\eeqn
The first term is analytic, while the second term has an infrared
renormalon located at $p/(2 b_0)$, which corresponds to a power correction
of the order of $1/Q^p$. Observe that for positive $b_0$ we have
found a definite prescription to bypass the IR pole.
However, this is not to be trusted, since for positive $b_0$ we should
interpret our results only as a power expansion in $\as$, as discussed
earlier.

In case the behaviour of $G(k^2)-G(0)$ is of the type
$k^p \log^n(k/Q)$ instead of a simple power,
it is easy to convince oneself that the $1/Q^p$ correction
will be enhanced by $n$ inverse powers of $\as$. In fact, the
inclusion of powers of $\log(k/Q)$ can be achieved from formula
(\ref{Ipdef}) by taking derivatives with respect to $p$.
Thus, since
the large order behaviour of the expansion of $I_p$ has the form
\beq
\propto \Gamma(n+1)\left(\frac{2b_0\as}{p}\right)^{n+1},
\eeq
taking a derivative with respect to $p$ we get the leading behaviour
\beq
\propto\frac{1}{p}\Gamma(n+2)\left(\frac{2b_0\as}{p}\right)^{n+1}
=\frac{1}{2 b_0\as}\Gamma(n^\prime+1)
\left(\frac{2b_0\as}{p}\right)^{n^\prime+1}
\eeq
with $n^\prime=n+1$, which corresponds to a $1/\as$ enhancement.

We have therefore found that in our approach the coefficient of the
power correction will depend upon the behaviour
of $S(\phi,k^2)$ for small $k^2$, which is to say, upon how the definition
of the shape variables for 4 partons goes to the 3--parton definition
in the collinear limit. This indicates that the coefficient of the power
correction cannot be simply factorized in terms of the three--body
definition of the shape variable, a fact that we will examine in more details
in the following sections.

\section{Details of the calculation}
We begin with some kinematical preliminaries.
%%%%%%%%%%%%%%%%%%%%%%%%
\begin{figure}[htb]
\centerline{\psfig{figure=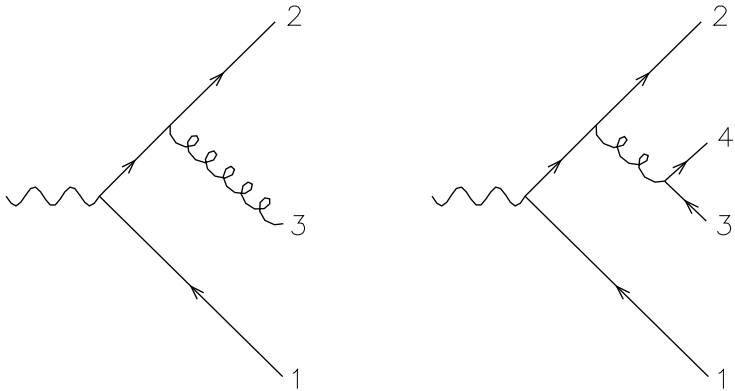,width=10cm,clip=}}
\caption{ \label{kinematics}
Labeling of external lines for three-- and four--parton processes.
}
\end{figure}
%%%%%%%%%%%%%%%%%%%%%%%%
Outgoing legs for the
three-- and four--parton process are given in fig.~\ref{kinematics}.
We will call $p_i$ the momenta of the outgoing legs,
$E_i$ their energies, and the invariants will be defined as
\beqn
 Q &=& \sqrt{\left(\sum p_i\right)^2} \nonumber \\
 s_{ij}&=&(p_i+p_j)^2,\quad y_{ij}=\frac{s_{ij}}{Q^2}\nonumber \\
 s_{ijk}&=&(p_i+p_j+p_k)^2,\quad y_{ijk}=\frac{s_{ijk}}{Q^2}.
\eeqn
For the 4--parton process we have
\beq
E_1=\frac{Q^2-s_{234}}{2Q}\,,\quad
E_2=\frac{Q^2-s_{134}}{2Q}\,.
\eeq
In the three-body case these simplify to
\beq
E_2=\frac{Q^2-s_{13}}{2Q}\quad\quad E_1=\frac{Q^2-s_{23}}{2Q}.
\eeq
The maximum value of $s_{34}$ for fixed $s_{134}$ and $s_{234}$ is reached
when $\vec{p}_1$ and $\vec{p}_2$ are parallel and opposite:
\beqn
%&& \sqrt{(E_1-E_2)^2+s_{34}}+E_1+E_2=Q
%\nonumber \\&&
%(E_1-E_2)^2+s_{34}=(Q-E_1-E_2)^2
%\nonumber \\&&
%\left(\frac{s_{234}-s_{134}}{2Q}\right)^2+s_{34}=
%\left(\frac{s_{134}+s_{234}}{2Q}\right)^2
%\nonumber \\&&
s_{34}=\frac{s_{234}s_{134}}{Q^2}.
\eeqn
Defining
\beq
x_1=\frac{2 E_1}{Q}=1-y_{234}\,,\quad
x_2=\frac{2 E_2}{Q}=1-y_{134}
\eeq
we have the constraint
\beq
s_{34}<Q^2(1-x_1)(1-x_2)\,.
\eeq
We follow here the calculation of ref.~\cite{ERT}, from which many
of the following results are taken. We work in $d=4-2\ep$ dimensions.
The three body cross section is given by the formula
\beqn
&& d\sigma^{(3)}=H\frac{\as}{2\pi}C_F\left(\frac{4\pi\mu^2}{Q^2}\right)^\ep
\frac{1}{\Gamma(1-\ep)} T(x_1,x_2)\nonumber \\&&
\theta(x_1+x_2-1)\left((1-x_1)(1-x_2)(x_1+x_2-1)\right)^{-\ep} \;\;dx_1\;dx_2.
\eeqn
The constant $H$ is the normalization of the Born 2 body cross section,
$\sigma^{(2)}=H$.
The four--parton cross section is
\beqn &&
d\sigma^{(4)}=H\frac{\as C_F}{2\pi}\left(\frac{4\pi\mu^2}{Q^2}\right)^\ep
\frac{1}{\Gamma(1-\ep)}
\left((1-x_1)(1-x_2)-y_{34}\right)^{-\ep}
(y_{34}+x_1+x_2-1)^{-\ep} \;
\nonumber \\&&
\theta\left((1-x_1)(1-x_2)-y_{34}\right)\;
\theta\left(y_{34}+x_1+x_2-1\right) \;\;dx_1\;dx_2
\frac{\as T_R}{2\pi}\left(\frac{4\pi\mu^2}{Q^2}\right)^\ep
\frac{1}{\Gamma(1-\ep)}
\nonumber \\&&
\left(z(1-z)\right)^{-\ep}\;dz\;\frac{1}{N_\theta'}\sin^{-2\ep}\theta'
d\theta'\;\;\frac{dy_{34}}{y_{34}^{1+\ep}}\;\;
T^{(4)}(x_1,x_2,y_{34},z,\theta^\prime)
\label{sigma4}
\eeqn
where
\beqn \label{TandV}
T^{(4)}(x_1,x_2,y_{34},z,\theta^\prime)&=&
T^{(4)}_{\rm coll}(x_1,x_2,z,\theta^\prime)+
V(x_1,x_2,y_{34},z,\theta^\prime)
\\
T^{(4)}_{\rm coll}(x_1,x_2,z,\theta^\prime)&=&
T(x_1,x_2)\;
\left(\frac{z^2+(1-z)^2-\ep}{1-\ep}\right)
\nonumber\\&&
-R(x_1,x_2)\;4z(1-z)
\left(2\cos\theta'-\frac{1}{1-\ep}\right)
\\
T(x_1,x_2)&=&\frac{x_1^2+x_2^2}{(1-x_1)(1-x_2)}+{\cal O}(\ep)
\\
R(x_1,x_2)&=&\frac{x_1+x_2-1}{(1-x_1)(1-x_2)}
\\
\int \frac{1}{N_\theta'}\sin^{-2\ep}\theta' d\theta'&=&1
\,.
\eeqn
The $V$ term (which can be extracted from ref.~\cite{ERT})
vanishes for $y_{34}\to 0$.
We have kept the four--parton phase space factorized into a three--parton
component, describing the production of a quark, an antiquark and a gluon,
and a two--parton term, corresponding to the decay of the virtual gluon
into a quark-antiquark pair.

In order to study the infrared renormalon, we must now include all
the vacuum polarizations in the three and four--parton processes.
First of all, we need a formula for the vacuum polarization in the
${\rm MS}$ scheme. We obtain
\beqn
\Pi_{\mu\nu}&=&\Pi(k^2)\;\left(g_{\mu\nu}\;k^2-k_\mu\,k_\nu\right)
\nonumber \\
\Pi(k^2)&=&-i\frac{\as T_R}{3\pi}\; \left(\frac{-k^2}{\mu^2}\right)^{-\ep}
\;\frac{N(\ep)}{\ep} \nonumber \\
N(\ep)&=&(4\pi)^\ep\,\Gamma(1+\ep)\frac{\Gamma^2(1-\ep)}{\Gamma(1-2\ep)}
\frac{1-\ep}{(1-\frac{2}{3}\ep)(1-2\ep)}\;
\stackrel{\ep\to 0}{\longrightarrow}\; 1\,.
\eeqn
It is now easy to show that insertion of all vacuum blobs into a gluon
line amounts to the following replacement
\beq
\as\; \to \;\frac{\as}
{1-\as b_0\left(\frac{-k^2}{\mu^2}\right)^{-\ep}\frac{N(\ep)}{\ep}}\,.
\eeq
The renormalizability of our theory implies that all divergences
are removed by a redefinition
\beq
\as \to Z\;\aren
\eeq
where in the MS scheme $Z$ is a power expansion in $\aren$, whose coefficients
contain only inverse powers of $\ep$. After renormalization, our
replacement rule will then become
\beq
\as\; \to\; \frac{\aren}
{Z^{-1}-\aren b_0\left(\frac{-k^2}{\mu^2}\right)^{-\ep}\frac{N(\ep)}{\ep}}
\eeq
so that we must have
\beq
Z^{-1}=1+\aren b_0\;\frac{1}{\ep}.
\eeq
Therefore, the resummation of all bubbles plus charge renormalization
amounts to the replacement
\beq
\as\; \to\; \frac{\as}
{1-\as b_0\left[\left(\frac{-k^2}{\mu^2}\right)^{-\ep}N(\ep)-1\right]
\frac{1}{\ep}}.
\eeq
We can now immediately write down the result for the three--parton
cross section including the effect of all bubble insertions.
In this case, in fact, the momentum flowing through the gluon propagator
is exactly zero. Indicating with a tilde the fully resummed cross section
we get
\beq
d \tilde{\sigma}^{(3)}\;=\;
 d \sigma^{(3)} \frac{1}{1+\as b_0 \frac{1}{\ep}}.
\eeq
Observe that in spite of the renormalization procedure we carried
out, poles in $\ep$ do remain, and they should in fact be interpreted
as infrared poles, that will ultimately cancel against analogous
contributions in the four--parton cross section.

The case of the four--parton cross section is more involved.
In this case one should remember that in the $\as^2$ factor
one power of $\as$ should be complex-conjugated. We then have
\beqn
d \tilde{\sigma}^{(4)} & = & d \sigma^{(4)} \frac{1}{1-\as\,b_0
\left(y\,e^{i\ep\pi}\,N(\ep)-1\right)\frac{1}{\ep}}
\;\;
\frac{1}{1-\as\,b_0\left(y\,e^{-i\ep\pi}\,N(\ep)-1\right)\frac{1}{\ep}}
\nonumber \\
&=& d \sigma^{(4)}
\frac{1}{\left[\left(1+\frac{\as b_0}{\ep}\right)\,\cos\ep\pi
-\frac{\as b_0}{\ep}N(\ep)\,y\right]^2+\left(1+\frac{\as b_0}{\ep}\right)^2
\sin^2\ep\pi}
\eeqn
where we have defined
\beq
y=\left(\frac{s_{34}}{\mu^2}\right)^{-\ep}.
\eeq
In order to make the infrared cancelation explicit, we will proceed as
follows. Let us call $G$ a generic infrared safe jet shape variable.
For our purposes, $G$ is characterized by two functions
\beq
G^{(3)}(x_1,x_2),\quad\quad G^{(4)}(x_1,x_2,y_{34},\theta',z)
\eeq
and infrared safety will imply that
\beq
\lim_{y_{34}\to 0}G^{(4)}(x_1,x_2,y_{34},\theta',z)=G^{(3)}(x_1,x_2).
\eeq
We are implicitly assuming that $G$ does not receive
contributions from the two--parton final state, and therefore it has
a power expansion that starts at order $\as$, which is the case for
all shape variables usually considered in $e^+e^-$ physics.
The value of $G$ in our model will be given by
\beq
G=\int d\;\tilde{\sigma}^{(3)} G^{(3)}(x_1,x_2)+\int d\;\tilde{\sigma}^{(4)} \;
G^{(4)}(x_1,x_2,y_{34},\theta',z).
\eeq
We now rewrite the above expression in the following form
\beq
G = G_{\rm virt}+G_{4}+G_{V}+G_{\rm coll}
\eeq
with
\beqn
G_{\rm virt} &=&\int d\;\tilde{\sigma}^{(3)} G^{(3)}(x_1,x_2)
\nonumber \\
G_{4} &=& \int d\;\tilde{\sigma}^{(4)}
 \;\Big[G^{(4)}(x_1,x_2,y_{34},\theta',z)-
F(x_1,x_2,y_{34})\,G^{(3)}(x_1,x_2)\Big]
\nonumber \\
G_{V}&=&\int d\;\tilde{\sigma}_{V}^{(4)}\; F(x_1,x_2,y_{34})\;G^{(3)}(x_1,x_2)
\nonumber \\
G_{\rm coll}&=&
\int d\;\tilde{\sigma}_{\rm coll}^{(4)}\; F(x_1,x_2,y_{34})\;G^{(3)}(x_1,x_2),
\eeqn
and
\beqn
F(x_1,x_2,y_{34})&=&\frac{((1-x_1)(1-x_2)-y_{34})^\ep\,(y_{34}+x_1+x_2-1)^\ep}
{((1-x_1)(1-x_2))^\ep\,(x_1+x_2-1)^\ep}
\theta(x_1+x_2-1)
\nonumber \\
d\tilde{\sigma}^{(4)}&=&d\tilde{\sigma}^{(4)}_{V}
+d\tilde{\sigma}^{(4)}_{\rm coll}.
\eeqn
The separation of $d\sigma^{(4)}$ into the collinear and $V$ term
is performed according to eq.~(\ref{TandV}).
The $F$ factor is chosen in order to simplify the $y_{34}$ integral
in the $G_{\rm coll}$ term.
The $G_4$ and $G_V$ terms are free of infrared singularities,
because the integrands vanish for $y_{34}\to 0$, so that
$\ep$ can be safely replaced by 0 in their expressions.
In ref.~\cite{Seymour} it was shown that after $\theta^\prime$ and
$z$ integration $G_V$ is of order $y_{34}$.
In $G_{\rm coll}$, the integration in $z$, $\theta'$ and $y_{34}$
can be performed, because $G^{(3)}$ does not depend upon these quantities.
The $z$ and $\theta'$ integrals are easily done.
The $R$ term vanishes after angular integration, and the $z$ integration
gives
\beq
\int dz \,(z(1-z))^{-\ep}\;\frac{z^2+(1-z)^2-\ep}{1-\ep}=
\frac{\Gamma^2(1-\ep)}{\Gamma(1-2\ep)}\frac{2(1-\ep)}{(1-2\ep)(3-2\ep)}
=\frac{2N(\ep)}{3\,(4\pi)^\ep\Gamma(1+\ep)}\,.
\eeq
Using the identity
\beq
\frac{d y_{34}}{y_{34}^{1+\ep}}=- \frac{dy}{\ep}\,\left(\frac{Q^2}{\mu^2}
\right)^\ep
\eeq
we compute the $y_{34}$ integral
\beqn &&
\int_0^{(1-x_1)(1-x_2)} \frac{d y_{34}}{y_{34}^{1+\ep}}
\frac{1}{\left[\left(1+\frac{\as b_0}{\ep}\right)\,\cos\ep\pi
-\frac{\as b_0}{\ep}N(\ep)\,y\right]^2+\left(1+\frac{\as b_0}{\ep}\right)^2
\sin^2\ep\pi} \nonumber \\ &&=
\frac{-\left(\frac{Q^2}{\mu^2}\right)^\ep}
{\as\,b_0\,N(\ep)\left(1+\frac{\as b_0}{\ep}\right)\sin\ep\pi}\;\;
\left. \arctan\frac{\left(1+\frac{\as b_0}{\ep}\right)\sin\ep\pi}
{\left(1+\frac{\as b_0}{\ep}\right)\cos\ep\pi-\frac{\as b_0}{\ep}\,N(\ep)y}
\right|^{(1-x_1)(1-x_2)}_0
\nonumber \\&&=
-\frac{1}{\as^2\,b_0^2\,\pi}\arctan\frac{\as\,b_0\,\pi}
{1+\as\,b_0\,\left(\log\frac{(1-x_1)(1-x_2)Q^2}{\mu^2}-N'(0)\right)}
\nonumber \\&&
+\left(\frac{Q^2}{\mu^2}\right)^\ep
\frac{\ep\pi}{\as\,b_0\,N(\ep)\left(1+\frac{\as b_0}{\ep}\right)\sin\ep\pi}
\eeqn
where we have set explicitly $\ep=0$ in the first term.
Using the identity
\beq
\frac{\ep\pi}{\sin\ep\pi\,\Gamma(1+\ep)\Gamma(1-\ep)}=1
\eeq
we write our integral as
\beqn &&
G_{\rm coll} =-\int\frac{\as C_F}{2\pi}
\theta\left(x_1+x_2-1\right) \;\;dx_1\;dx_2
\nonumber \\&&
\times \frac{\as T_R}{3\pi} T(x_1,x_2)\;G^{(3)}(x_1,x_2)
\frac{1}{\as^2\,b_0^2\,\pi}\arctan\frac{\aseff\,b_0\,\pi}
{1+\aseff\,b_0\,\log\frac{(1-x_1)(1-x_2)Q^2}{\mu^2}}
\nonumber \\&&
+ \int H\frac{\as C_F}{2\pi}\left(\frac{4\pi\mu^2}{Q^2}\right)^\ep
\frac{1}{\Gamma(1-\ep)}
\theta\left(x_1+x_2-1\right)[(1-x_1)(1-x_2)(x_1+x_2-1)]^{-\ep}
\nonumber \\&&
\times \frac{\as T_R}{3\pi}
\frac{1}{\as\,b_0\,\left(1+\frac{\as b_0}{\ep}\right)}
T(x_1,x_2)\;G^{(3)}(x_1,x_2) \;\;dx_1\;dx_2
\label{collint}
\eeqn
where we have defined according with ref.~\cite{aseff}
\beq
\frac{1}{\aseff}=\frac{1}{\as}-N^\prime(0).
\eeq
Remembering the identity for $b_0$ we see that the second integral
in the above equation cancels exactly $G_{\rm virt}$,
so that all infrared divergences cancel, and we can write
\beq
G=G_4+G_V+G_S\,,\quad\quad G_S=G_{\rm virt}+G_{\rm coll}\,.
\eeq
The $G_S$ term is given by the first term of eq.~(\ref{collint}), and
the $S$ suffix stands for ``Sudakov'', since (in some sense) this
term comes from the incomplete cancelation between the real and
virtual diagrams, induced by the running of $\as$. It can be written as
\beq \label{SudakovTerm}
G_S=\frac{C_F}{2\pi}\int_{x_1+x_2<1}
\;\;dx_1\;dx_2\;
 T(x_1,x_2)\;G^{(3)}(x_1,x_2)
\frac{\arctan(\aseff(k_\perp^2)\,b_0\,\pi)}{b_0\,\pi}
\eeq
where
\beq
k_\perp^2=(1-x_1)(1-x_2)\;Q^2\,.
\eeq

\section{The Sudakov term}
A term of the form of $G_S$ appeared first in ref.~\cite{Webber}, and
was there used to parameterize the $1/Q$ correction to shape variables
like the average value of $1-t$, where $t$ is the thrust.
In ref.~\cite{Akhoury} it was argued that the
power corrections of the order $1/Q$ factorize in the form of
eq.~(\ref{SudakovTerm}) for a generic shape variable, as well as for other
processes.
In our calculation, this term, before
the $x_1,x_2$ integration, does not have any renormalon, since it is
an analytic function of $\as$ near the origin.
When we integrate over $x_1,x_2$, and approach the singular two--jet
region, non--analytic behaviour may arise. This is due to the fact
that we are integrating over the $x_1,x_2$ values where
\beq
1/\aseff(k_\perp^2)=
1+\aseff b_0\log\frac{k_\perp^2}{\mu^2}\;\to\; 0\,.
\eeq
Our formula differs from the one of ref.~\cite{Webber}
only by the replacement
\beq
\aseff(q^2)\to \frac{1}{b_0\pi}\arctan(\aseff(q^2) b_0 \pi)\,.
\eeq
In our case, however, there is really no singularity when we integrate
over the Landau pole, since the arctangent is a bounded function.
Let us compute
the contribution of the $G_S$ term to the average value of
$1-t$. For a three massless body system, thrust is simply
\beq
t=\max(x_1,x_2,2-x_1-x_2).
\eeq
The leading contribution to $1-t$ comes from the region
where $x_1$ and $x_2$ are very near 1, so
\beqn
\langle 1-t \rangle_S &=& \frac{C_F}{2\pi}2\int_{x_1>x_2} d x_1 dx_2
\; \frac{2(1-x_1)}{(1-x_1)(1-x_2)}
\frac{\arctan(\aseff(k_\perp^2)b_0\pi)}{b_0\pi}
\nonumber \\
&=& \frac{2 C_F}{\pi}\int_0^1 \frac{d y}{\sqrt{y}}\;
\frac{\arctan(\aseff(y Q^2) b_0\pi)}{b_0\pi}\,.
\eeqn
In order to evidentiate the structure of the infrared renormalon in the
above formula, we compute the integral
\beq
I=\int_0^1 \frac{d y}{y}\;y^\frac{p}{2}
\frac{\arctan(\aseff(y Q^2) b_0\pi)}{b_0\pi}
\eeq
for arbitrary $p>0$. We integrate by parts, and obtain
\beqn
I&=&\frac{2\arctan(\aseff(Q) b_0\pi)}{p\,b_0\,\pi}
+\frac{2}{p}\int_0^1 \frac{dy}{y}\,y^{\frac{p}{2}} \frac{b_0}{
\frac{1}{\aseff^2(yQ^2)}+(b_0\pi)^2}
\nonumber \\
&=&\frac{2\arctan(\aseff(Q) b_0\pi)}{p\,b_0\,\pi}+\frac{2b_0}{p}\;I_p\,.
\eeqn
The first term is analytic in $\aseff$ near the origin, while the second term,
given by eq.~(\ref{IpDef}) has an infrared renormalon located
at $z=p/(2 b_0)$, which corresponds to a $1/Q^p$ power correction.
For the case of $\langle 1-t \rangle$ we found therefore a $1/Q$
correction.

\section{The four--parton integral}
The terms $G_4$ and $G_V$ cannot easily be done analytically,
because they depend in an intricate way on the four--parton
phase space. Observe that
\beqn
G_{4}+G_{V}&=& \int \Big[\; d\,\tilde{\sigma}^{(4)}
 \; G^{(4)}(x_1,x_2,y_{34},\theta',z)-\theta(x_1+x_2-1)\;
 d\,\tilde{\sigma}^{(4)}_{\rm coll}\,G^{(3)}(x_1,x_2)\;\Big]
\nonumber \\
&=& \int d y\; G_{4V}(y) \frac{\aseff^2}{\left(1+\aseff
b_0\log\frac{yQ^2}{\mu^2} \right)^2+(\aseff b_0\pi)^2}\,,
\eeqn
where we have defined
\beqn
G_{4V}(y)&=& \frac{1}{\aseff^2}\int \delta(y-y_{34}) \Big[\; d\,\sigma^{(4)}
 \; G^{(4)}(x_1,x_2,y_{34},\theta',z)
\nonumber \\&&
-\theta(x_1+x_2-1)\;
 d\,\sigma^{(4)}_{\rm coll}\,G^{(3)}(x_1,x_2)\;\Big]\,.
\eeqn
It is clear that the small $y$ behaviour of $G_{4V}(y)$ controls
the power correction due to the IR renormalon. In particular,
if
\beq
G_{4V}(y)\stackrel{y\to 0}{\longrightarrow} A\; \frac{y^{p/2}}{y}
\eeq
the position of the renormalon will be at $p/(2b_0)$, corresponding to
a power correction $1/Q^{p}$.

We computed $G_{4V}$ numerically for $y=10^{-j}$, $j=1,\ldots 6$, for the
following shape variables: $\langle 1-t \rangle $, $\langle
\theta(0.8-t)\rangle$, $\langle m_H^2\rangle$, where $m_H^2$ is the
heavy jet mass-squared according to the thrust definition, $\langle
\theta(m_H^2-0.1)\rangle$, $\langle \theta(o-0.1)\rangle$, $\langle
\theta(c-0.2)\rangle$, and for the
weighted average of the energy-energy correlation away from the
back-to-back region
\beq
EEC_{\rm cut}=
 \int_{-0.5}^{0.5}
EEC(\cos \theta)\; \sin^2\theta\;d \cos\theta.
\eeq
For the exact definition of these quantities, see ref.~\cite{Yellow}.
The results are given in table~\ref{4partres}.
For each value $y=10^{-j}$,
we also give the power $p$ that is obtained by fitting $G_{4V}(y)$
in the two points $10^{-(j-1)}$ and $10^{-j}$ with a function
proportional to $y^{p/2}/y$.
%%%%%%%%%%%%%%%%%%%%%%%%%%%%%%%%%%%%%%%%%%%%%%%%%%%%%%%%%%%%%%%%%%%%%%%%%%%%
\begin{table}
\begin{center}
\begin{tabular}{|l||c|c|c|c|c|c|} \hline
$y$&$10^{-1}$&$10^{-2}$&$10^{-3}$&$10^{-4}$&$10^{-5}$&$10^{-6}$
\\ \hline
$\langle 1-t \rangle $&
.2365(11)&-2.146(7)&-12.21(3)&-46.34(15)&-159.2(6)&-522(3)\\\hline
$p$& *&*&.490(4)&.842(4)&.928(5)&.969(6)\\\hline
$t<0.8$&
.611(3)&-.566(13)&1.435(10)&2.341(10)&2.469(10)&2.477(10)\\\hline
$p$& *&*&*&1.575(7)&1.954(5)&1.997(5)\\\hline
$\langle m_H^2 \rangle$&
.0437(11)&-4.595(12)&-25.31(7)&-109.6(4)&-435.8(19)&-1639(9)\\\hline
$p$& *&*&.518(3)&.727(4)&.801(5)&.849(6)\\\hline
$m^2_H>0.1$&
.384(2)&-30.22(7)&-62.8(3)&-66.0(5)&-13.60(15)&-.035(10)\\\hline
$p$& *&*&1.365(4)&1.956(7)&3.372(11)&7.2(2)\\\hline
$o>0.1$&
.252(4)&-119.6(3)&-1856(5)&-5250(30)&-14610(140)&-44600(800)\\\hline
$p$& *&*&-.381(3)&1.098(5)&1.110(9)&1.031(17)\\\hline
$c>0.2$&
.934(3)&-21.11(11)&-110.5(10)&-129(3)&-145(11)&-150(30)\\\hline
$p$& *&*&.563(9)&1.86(2)&1.90(7)&2.0(2)\\\hline
$ EEC_{\rm cut}$&
.4150(18)&-2.12(2)&-13.9(2)&-51.8(13)&-180(8)&-630(40)\\\hline
$p$& *&*&.371(16)&.85(3)&.92(4)&.92(7)\\\hline
\end{tabular}
\caption[tbh]{  \label{4partres}
Results for $G_{4V}(y)$ for various shape variables, for
$y=10^{-1},\ldots 10^{-6}$. The line marked $p$,
in the column corresponding to $y=10^{-k}$, is the exponent
one would obtain from the above table by fitting the pair of numbers
on the line above,
corresponding to  $y=10^{-k+1}$ and  $y=10^{-k}$ with the form
$y^{p/2}/y$.
}
\end{center}
\end{table}
%%%%%%%%%%%%%%%%%%%%%%%%%%%%%%%%%%%%%%%%%%%%%%%%%%%%%%%%%%%%%%%%%%%%%%%%%%%
As anticipated in the previous section, we can see from the table that
the four--parton contribution
can give $1/Q$ power suppressed corrections to quantities like
the average value of $1-t$.

It is easy to identify regions of integration that give such type
of contributions. Consider for example the region
\beq
\sqrt{y_{34}}<(1-x_1)<\sqrt{y_{34}} (1+\eta)
\,\quad\quad
\sqrt{y_{34}}<(1-x_2)<\sqrt{y_{34}} (1+\eta)
\eeq
for $\eta\ll 1$ and independent of $y_{34}$. In this configuration,
$1-t$ is always of order $\sqrt{y_{34}}$.
For small $y_{34}$ the emitted gluon
is soft, so that the amplitude factorizes in terms of the
three body amplitude in the soft limit, and a term depending
upon the orientation of partons 3 and 4. The three body amplitude gives
an integral of the form
\beq
\int_{\sqrt{y_{34}}}^{\sqrt{y_{34}} (1+\eta)}
 \frac{d(1-x_1)}{1-x_1} \frac{d(1-x_2)}{1-x_2}=\eta^2+{\cal O}(\eta^3),
\eeq
independent of $y_{34}$.
Next, we have to weight the amplitude with $1-t$, which is of
order $\sqrt{y_{34}}$, and integrate in $d\,y_{34}/y_{34}$.
This corresponds to $G_{4V}\propto \sqrt{y_{34}}/y_{34}$,
which, as we have seen, yields a $1/Q$
correction.

One may wonder whether  one may still recover a factorized form,
similar to the term $G_S$, by a suitable redefinition of the
effective coupling.
Looking at the quantity
$\langle m_H^2 \rangle $ we see that this is not the case.
In fact, the invariant mass of the heavy jet is equivalent at the
3--parton level to thrust, namely we have $1-t=m_H^2/Q^2$.
This identity is no longer valid at the
4--parton level. For example, in the soft configuration we have just
considered, where partons 1 and 2 are back--to--back,
and so are partons 3 and 4, it is a simple exercise
to show that the relation is instead $1-t=2 m_H^2/Q^2$. Therefore, any
three--body factorization formula would fail in this simple
case. Shape variables that are identical at the
3--parton level, but differ at the 4--parton level,
have different coefficients for the leading power correction.

Let us now focus on shape variables that depend upon final state
configurations that are far from the two jet region, for which
the Sudakov term does not provide a leading $1/Q$
power correction. For some of these variables, e.g.
$\langle\theta(0.8-t)\rangle$, $\langle\theta(m_H^2-0.1)\rangle$,
$\langle\theta(c-0.2)\rangle$, we see no evidence for
power corrections of the form $1/Q$, but instead we find
a $1/Q^2$ correction (in the case of $\langle\theta(m_H^2-0.1)\rangle$,
the $1/Q^2$ term has a rather small coefficient, so this behaviour does
not become apparent until $y\ltap10^{-8},$ where $G_{4V}(y)$ becomes
constant at +1.43, indicating $p=2$).
For $\langle\theta(o-0.1)\rangle$ we instead
observe a $1/Q$ type of correction. A leading $1/Q$ correction
is also observed for $EEC_{\rm cut}$, in spite of the cut that avoids the
back--to--back region. As we will see shortly, this is due to the fact that
$EEC$ receives contributions from configurations near the two jet region
also for angles far away from the back--to--back configuration.

\begin{figure}[htb]
\centerline{\psfig{figure=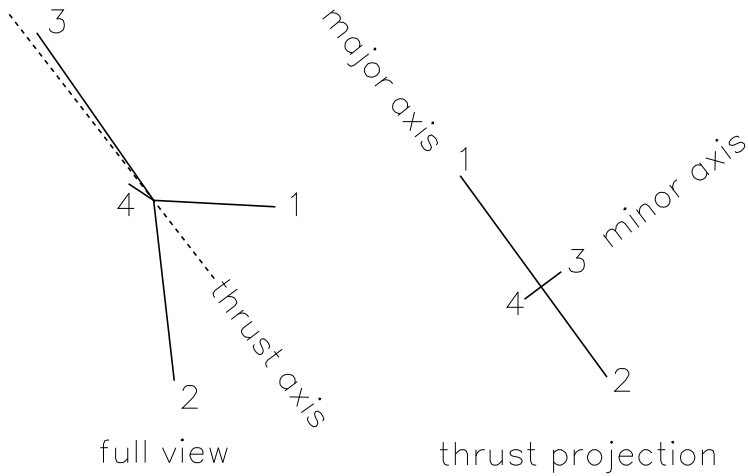,width=10cm,clip=}}
\caption{ \label{oblateness}
Oblateness in the collinear limit.
}
\end{figure}

These findings can be easily justified by examining the singular integration
region for various quantities.
First of all we will consider thrust.
Let us look at parton configurations near the collinear limit.
By kinematical reasoning one can convince oneself that the thrust axis
is either along parton 1, parton 2, or along the sum of partons 3 and 4,
and the thrust is given by $x_1$, $x_2$, or
$(2-x_1-x_2)-2y_{34}/t+{\cal{O}}(y_{34}^2)$ respectively.
In all cases, it differs from the thrust
of the corresponding configuration with $y_{34}=0$ by terms
of the order of $y_{34}$ or less. This behaviour gives rise to a
$1/Q^2$ power correction.
In the case of oblateness, we can instead identify
a region where a $\sqrt{y_{34}}$ behaviour arises, leading
to a $1/Q$ power correction.
One such configuration is depicted in Figure~\ref{oblateness}.
%%%%%%%%%%%%%%%%%%%%%%%%
%%%%%%%%%%%%%%%%%%%%%%%%
One projects the event onto the plane orthogonal to the thrust
axis, and then oblateness is defined as the difference between
the major and minor axis. For the particular configuration shown
in the figure, oblateness is just the difference between the distance
1--2 (called $f$--major) and the distance 3--4 (called $f$--minor)
in the projected event.
It is easy to convince oneself that the 3--4 distance is proportional
to $\sqrt{y_{34}}$. In the limit of $y_{34}\to 0$ oblateness behaves
therefore like $\sqrt{y_{34}}$, which generates a $1/Q$ correction.

The $1/Q$ correction in the case of the energy--energy correlation has
instead a very different origin.
One such correction arises from the Sudakov term.
In the 3--body configuration, when parton 3 is soft,
the Sudakov term for the energy--energy correlation is
\beq
  \sin^2\theta \; EEC(\cos\theta) = 8\,\frac{C_F}{2\pi}\frac1{\sin\theta}
  \int_0^{\frac12Q\sin\theta} \frac{d\kp}{Q} \;
  \frac{\arctan(\ask b_0\pi)}{b_0\pi}.
\eeq
The integral over $\kp$ generates a $1/Q$ correction.
Other contributions come from the $G_{4V}$ term.
Consider some weighted average of the energy--energy correlation,
with a weight $f(\theta)$
in an angular interval that does not include the back--to--back region.
Then, for example, the contributions coming from partons 1 and 3
would be
\beq
\int d\sigma_3\; f(\theta)\; E_1\, E_3\,.
\eeq
If parton 3 (the gluon) splits into a quark--antiquark pair,
carrying fractions $z$ and $1-z$ of parton 3's momentum, and having
an opening angle $\omega$. Also assume for simplicity that the
splitting takes place in the 1--3 plane.
The contribution in the collinear limit is then
\beqn &&
\int d\sigma_4\; E_1\,E_3\;\left(
f(\theta+z\omega)(1-z)+f(\theta-(1-z)\omega)z\right)
\nonumber \\ &&
\approx \int  d\sigma_4\; E_1\,E_3\;\left(f(\theta)
+\frac{1}{2}\;f ^{\prime\prime}(\theta)\omega^2\,z(1-z) \right)
\nonumber \\ &&
= \int d\sigma_4\; E_1\,E_3\;\left(f(\theta)
+\frac{1}{2}\;f^{\prime\prime}(\theta)s_{34}/E_3^2 \right),
\label{EECexample}
\eeqn
which seems to give rise to a $1/Q^2$ power correction.
This is in fact not the case, since we integrate over the region where
$E_3$ becomes small.
In this region the cross section behaves as $d E_3/E_3$, so the $E_3$ integral
in eq.~(\ref{EECexample}) yields
\beq
\int_{\sqrt{s_{34}}} \frac{d E_3}{E_3^2}\propto \frac{1}{\sqrt{s_{34}}}\,,
\eeq
and we see that in this way a $1/Q$ power correction does arise.
Therefore, the energy--energy correlation receives $1/Q$ power corrections
for all values of the angle, coming from the region near the two--jet
limit.  It might appear that if we set $f(\theta)={}$constant, this $1/Q$
correction would be zero.  However there would still remain
$\theta$-function weights defining the edges of the integration region,
which would give equivalent $1/Q$ terms.

Unlike the case of oblateness, we see that for the EEC the correction arises
because the kinematic region
near the two--jet configuration contributes for all values of the angle.
All other commonly considered shape variables
depend instead upon the three jet region for intermediate values of the
shape parameter.

\section{Higher order terms}
We have seen from the previous section that, except for special cases
like oblateness, $1/Q$ corrections arise from
configurations with a soft gluon emission, where the gluon
virtuality is of the order of its energy, followed by the decay of the virtual
gluon into massless partons. This process cannot occur
away from the two jet region at leading $n_f$, but it can
certainly arise at subleading $n_f$. We may imagine adding to a
three--jet, $q\bar{q} g$ configuration a soft, off--shell gluon (i.e.
with energy of the same order as its virtuality)
decaying into a massless parton pair.
It is difficult to imagine any shape variable
that will not receive $1/Q$ corrections from this kind of process.
In fact, shape variables are typically linear in the parton momenta,
as dictated by the requirement of insensitivity to
collinear splitting. The production of a soft, off--shell gluon
reduces linearly the energy available to all the other partons,
which in general may affect the shape variable linearly in the gluon energy.
Since the cross section for soft gluon emission
has the characteristic behaviour $dE_g/E_g$,
and the emission coupling will be evaluated at the virtuality
of the gluon (assumed to be of the same order as
$E_g$) it follows that $1/Q$ corrections
are present. We have not, of course, rigorously proven this
fact. Needless to say, if shape variables that never develop $1/Q$
corrections were found,
their importance for the determination of $\as$ would be enormous.

Let us therefore assume, for a moment, the pessimistic
(and perhaps realistic) view that shape variables always develop
a $1/Q$ correction at some order in perturbation theory.
Let us consider, for example, thrust with a cut $t<0.8$,
so that we are always in the three jet region. According to the
above argument an extra soft gluon emission will generate
a correction of order $1/Q$.
It seems plausible however that
the hard real emission contributes a factor of $\as(Q^2)$,
such that the overall correction is of the
order of $\as(Q)/Q$. This would again be a very important
fact. It would tell us that some shape variables are indeed better
than others, in the sense that their $1/Q$ power correction
carries an extra $\as(Q)$ suppression.

It may also be possible
that the $1/Q$ suppression will turn out to be enhanced by a power
of $\log(Q/\Lambda)$, which would compensate the $\as$ suppression.
This could be produced, for example,
by a 5--parton term that behaved as  $\sqrt{y_{45}}\log y_{45}$
when particles 4 and 5 become collinear.
Whether these logarithmic enhancements are present or not is a matter
that ought to be clarified with further studies. In the present
work, we simply remark that it is conceivable that one may find shape
variables in which the enhancement is not present, and that therefore
do have a $\as$ suppression of the $1/Q$ power corrections.

\section{Conclusions}
In the present work, we have proven that even in the simple model
of QCD at large $n_f$, shape variables in $e^+e^-$ annihilation
show remarkably different properties with regard to power corrections
originating from infrared renormalons. In particular, we have shown that
in the large $n_f$ limit, variables like $\langle 1-t \rangle$,
$\langle m_H^2 \rangle$, and the $EEC$ for any value of the angle,
develop a $1/Q$ correction, while thrust and the $c$ parameter
do not develop any $1/Q$ correction in the region where the two
jet configuration does not contribute. Another remarkable result
is that oblateness develops a $1/Q$ correction even away from the
two jet region.

We compare our findings with the results of refs.~\cite{Webber} and
\cite{Akhoury}.
We recover a correction
term with the factorized form proposed there,
but we also find an extra correction that spoils
factorization, since it specifically depends upon the 4--parton definition of
the shape variable. Thus, two shape variables that have the
same 3--body expression, but differ at the 4--body level, will have
different power corrections.
The discrepancy with the authors of ref.~\cite{Webber} can be tracked back
to the fact that they assume to some extent the validity of the perturbative
expansion even when the scale of $\as$ is very low,
a fact that does not take place in our calculation in the leading
$n_f$ limit.

We argue that even shape variables that do not develop a $1/Q$ correction
at the leading large $n_f$ level,
may develop one at subleading level,
and therefore in the full QCD.
We conjecture that the leading power correction to shape variables
will have in general the form $\alpha^n_{\rm S}(Q)/Q$,
and one may classify shape variables according
to the value of $n$.
It may therefore be possible to find a class of shape variables
with leading power correction of the form $\as(Q)/Q$. With these
shape variables, the influence of non--perturbative effects
upon the determination of $\as$ would be truly negligible
at LEP energies.

\section*{Acknowledgements}
We wish to thank G. Altarelli, Yu.L. Dokshitzer
and B.R. Webber for useful discussions.

\end{document}